\documentclass{article}
\usepackage{amsmath,amssymb,graphicx}

\newcommand{\as}{\alpha_s}
\newcommand{\J}{\widetilde{J}}

\title{Some topics in heavy-quark physics}
\author{A.G.~Grozin\\
Budker Institute of Nuclear Physics}
\date{}

\begin{document}

\maketitle

\begin{abstract}
Some topics which can be easily explained to undergraduate students
are presented, with elementary derivations.
For a more systematic treatment of heavy-quark physics,
see the textbook~\cite{MW:00}.
\end{abstract}

\section{Introduction}

$B$ meson is the hydrogen atom of Quantum Chromodynamics (QCD),
the simplest nontrivial hadron.
In the leading approximation, the $b$ quark in it
just seats at rest at the origin and creates chromoelectric field.
Light constituents (gluons, light quarks and antiquarks)
move in this external field.
Their motion is relativistic;
the number of gluons and light quark-antiquark pairs in this light cloud
is undetermined and varying.
Therefore, there are no reasons to expect that a nonrelativistic
potential quark model describes $B$ meson well enough
(in contrast to the $\Upsilon$ meson, where the nonrelativistic
two-particle picture gives a good starting point).

Similarly, $\Lambda_b$ baryon can be called the helium atom of QCD.
Unlike in atomic physics, where the hydrogen atom is much
simpler than helium,
$B$ and $\Lambda_b$ are equally difficult.
Both have a light cloud with a variable number of relativistic particles.
The size of this cloud is the confinement radius $1/\Lambda_{\text{QCD}}$;
its properties are determined by the large-distance nonperturbative QCD.

The analogy with atomic physics can tell us a lot about hadrons
with a heavy quark.
The usual hydrogen and tritium have identical chemical properties,
despite the fact that the tritium nucleus is 3 times heavier than the proton.
Both nuclei create identical electric fields, and both stay at rest.
Similarly, $D$ and $B$ mesons have identical ``hadro-chemical'' properties,
despite the fact that $b$ quark is 3 times heavier than $c$.

The proton magnetic moment is of the order of the nuclear magneton $e/(2m_p)$,
and is much smaller than the electron magnetic moment $e/(2m_e)$.
Therefore, the energy difference between the states of the hydrogen atom
with the total spins 0 and 1 (hyperfine splitting) is small
(of the order $m_e/m_p$ times the fine structure).
Similarly, the $b$ quark chromomagnetic moment is proportional to $1/m_b$
by dimensionality, and the hyperfine splitting between $B$ and $B^*$ mesons
is small (proportional to $1/m_b$).
Unlike in atomic physics, both ``gross'' structure intervals
and fine structure intervals are just some numbers times $\Lambda_{\text{QCD}}$,
because light components are relativistic
(a practical success of constituent quark models shows that
these dimensionless numbers for fine splittings can be rather small,
but they contain no small parameter).

In the limit $m\to\infty$,
the heavy quark spin does not interact with gluon field.
Therefore, it may be rotated at will, without changing physics.
Such rotations can transform $B$ and $B^*$ into each other;
they are degenerate and have identical properties in this limit.
This heavy quark spin symmetry yields many useful relations
among heavy-hadron form factors~\cite{IW:90}.
Not only the orientation, but also the magnitude of the heavy quark spin
is irrelevant in the infinite mass limit. 
We can switch off the heavy quark spin, making it spinless,
without affecting physics.
This trick considerably simplifies counting independent form factors,
and we shall use it often.
Or, if we wish, we can make the heavy quark to have spin 1;
it does not matter.

This leads to a supersymmetry group called
the superflavour symmetry~\cite{GW:90,Ca:91}.
It can be used to predict properties of hadrons containing
a scalar or vector heavy quark.
Such quarks exist in some extensions of the Standard Model
(for example, supersymmetric or composite extensions).
This idea can also be applied to baryons with two heavy quarks.
They form a small-size bound state
(with the radius of order $1/(m\as)$)
which has spin 0 or 1 and is antitriplet in colour.
Therefore, these baryons are similar to mesons with a heavy antiquark
having spin 0 or 1.
Accuracy of this picture cannot be high,
because even the radius of the $bb$ diquark is only a few times
smaller than the confinement radius.

\section{Mesons with a heavy quark}

Let's consider mesons with the quark contents $\bar{Q}q$,
where $Q$ is a heavy quark with mass $m$ ($c$ or $b$),
and $q$ is a light quark ($u$, $d$, or $s$).
As discussed above, the heavy quark spin is inessential
in the limit $m\to\infty$, and may be switched off.
In the world with a scalar heavy antiquark,
$S$-wave mesons have angular momentum and parity $j^P=\frac{1}{2}^+$;
$P$-wave mesons have $j^P=\frac{1}{2}^-$ and $\frac{3}{2}^-$.
The energy difference between these two $P$-wave states (fine splitting)
is a constant times $\Lambda_{\text{QCD}}$ at $m\to\infty$,
just like splittings between these $P$-wave states and the ground state;
however, this constant is likely to be small.

In our real world, the heavy antiquark $\bar{Q}$
has spin and parity $s_Q^P=\frac{1}{2}^-$.
The quantum numbers of the above paragraph are those
of the light fields' cloud of a meson.
Adding the heavy antiquark spin, we obtain, in the limit $m\to\infty$,
a degenerate doublet of $S$-wave mesons with spin and parity
$s^P=0^-$ and $1^-$,
and two degenerate doublets of $P$-wave mesons,
one with $s^P=0^+$ and $1^+$, and the other with $s^P=1^+$ and $2^+$.
At a large but finite heavy-quark mass $m$,
these doublets are not exactly degenerate.
Hyperfine splittings, equal to to some dimensionless numbers
times $\Lambda_{\text{QCD}}^2/m$, appear.
It is natural to expect that hyperfine splittings in $P$-wave mesons
are less than in the ground-state $S$-wave doublet,
because the characteristic distance between the quarks is larger
in the $P$-wave case.
Note that the $1^+$ mesons from the different doublets
don't differ from each other by any exactly conserving quantum numbers,
and hence can mix.
They differ by the angular momenta of the light fields,
which is conserved up to $1/m$ corrections;
therefore, the mixing angle should be
of order of $\Lambda_{\text{QCD}}/m$.

Mesons with $q=u$ and $d$ form isodoublets;
together with isosinglets with $q=s$, they form $SU(3)$ triplets.

\begin{figure}[ht]
\begin{center}
\begin{picture}(60,50)
\put(35.5,29){\makebox(0,0){\includegraphics{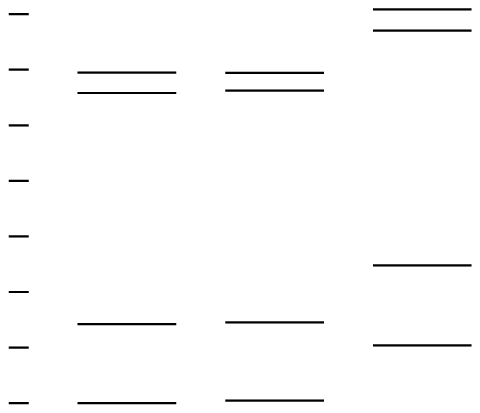}}}
\put(10,9){\makebox(0,0)[r]{0}}
\put(10,20.29){\makebox(0,0)[r]{200}}
\put(10,31.58){\makebox(0,0)[r]{400}}
\put(10,42.88){\makebox(0,0)[r]{600}}
\put(39,4){\makebox(0,0){$\bar{D}{}^0=\bar{c}u\hspace{3mm}
\bar{D}{}^-=\bar{c}d\hspace{3mm}
\bar{D}{}_s^-=\bar{c}s$}}
\put(24,11){\makebox(0,0){$\bar{D}\hspace{3mm}0^-$}}
\put(24,19.02){\makebox(0,0){$\bar{D}{}^*\hspace{3mm}1^-$}}
\put(24,38.51){\makebox(0,0){$\bar{D}{}_1\hspace{3mm}1^+$}}
\put(24,44.58){\makebox(0,0){$\bar{D}{}_2\hspace{3mm}2^+$}}
\end{picture}
\end{center}
\caption{Mesons with $\bar{c}$}
\label{cmes}
\end{figure}

Experimentally observed~\cite{PDG:00} mesons containing
the $\bar{c}$ antiquark are shown in Fig.~\ref{cmes}.
The energy scale at the left is in MeV, relative to the lowest mass meson.
The mesons $\bar{D}_1$ and $\bar{D}_2$ form the doublet
with the light fields' quantum numbers $j^P=\frac{3}{2}^+$.
The second $P$-wave doublet is suspiciously absent.
It should be close to the $\frac{3}{2}^+$ one;
it is not more difficult to produce these mesons than
the $\frac{3}{2}^+$ ones.
The problem is that they are too wide,
and cannot be cleanly separated from the continuum.

\begin{figure}[ht]
\begin{center}
\begin{picture}(60,50)
\put(35.5,29){\makebox(0,0){\includegraphics{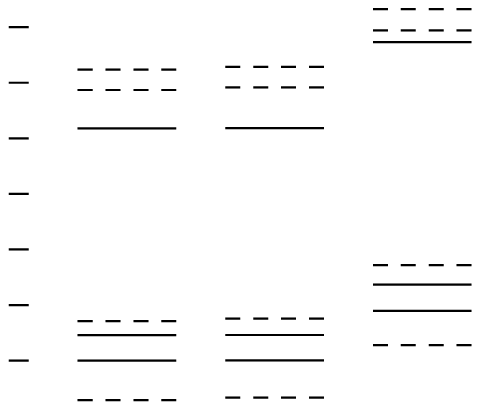}}}
\put(10,13.02){\makebox(0,0)[r]{0}}
\put(10,24.31){\makebox(0,0)[r]{200}}
\put(10,35.60){\makebox(0,0)[r]{400}}
\put(10,46.89){\makebox(0,0)[r]{600}}
\put(39,4){\makebox(0,0){$B^+=\bar{b}u\hspace{3mm}
B^0=\bar{b}d\hspace{3mm}
B_s^0=\bar{b}s$}}
\put(24,11.02){\makebox(0,0){$B\hspace{3mm}0^-$}}
\put(24,17.60){\makebox(0,0){$B^*\hspace{3mm}1^-$}}
\put(24,34.61){\makebox(0,0){$B_{1,2}\hspace{2mm}1^+,2^+$}}
\end{picture}
\end{center}
\caption{Mesons with $\bar{b}$}
\label{bmes}
\end{figure}

In the leading approximation,
the spectrum of $\bar{b}$ containing mesons
is obtained from the spectrum of $\bar{c}$ containing mesons
simply by the shift by $m_b-m_c$.
Experimentally observed mesons containing the $\bar{b}$ antiquark
are shown in Fig.~\ref{bmes}.
The spectrum of $\bar{c}$ containing mesons is shown
by dashed lines for comparison.
It is positioned in such a way that the weighted average energies
of the ground-state doublets coincide,
where the $1^-$ meson has weight 3 and the $0^-$ meson has weight 1.
The states $B_1$ and $B_2$ are not resolved,
and are shown by a single line.
Experimentally, it is difficult to measure their masses exactly enough.
The hyperfine splitting of the ground-state doublet
is smaller for $B$ mesons than for $D$ mesons, as expected.

\section{Baryons with a heavy quark}

In $S$-wave $Qqq$ baryons,
the light quark spins can add giving $j^P=0^+$ or $1^+$.
In the first case their spin wave function is antisymmetric;
the Fermi statistics and the antisymmetry in colour
require an antisymmetric flavour wave function.
Hence the light quarks must be different;
if they are $u$, $d$, then their isospin is $I=0$.
With the heavy quark spin switched off,
this gives the $0^+$ baryon $\Lambda_Q$ with $I=0$.
If one of the light quarks is $s$,
we have the isodoublet $\Xi_Q$,
which forms a $SU(3)$ antitriplet together with $\Lambda_Q$.
With the heavy quark spin switched on,
these baryons have $s^P=\frac{1}{2}^+$.
In the $1^+$ case, the flavour wave function is symmetric.
If the light quarks are $u$, $d$, then their isospin is $I=1$.
This gives the $1^+$ isotriplet $\Sigma_Q$;
with one $s$ quark -- the isodoublet $\Xi'_Q$;
with two $s$ quarks -- the isosinglet $\Omega_Q$.
Together they form the $SU(3)$ sextet.
With the heavy quark spin switched on,
we obtain the degenerate doublets with $s^P=\frac{1}{2}^+$, $\frac{3}{2}^+$:
$\Sigma_Q$, $\Sigma^*_Q$; $\Xi'_Q$, $\Xi^*_Q$; $\Omega_Q$, $\Omega^*_Q$.
The hyperfine splittings in these doublets
are of the order of $\Lambda_{\text{QCD}}^2/m$.
Mixing between $\Xi_Q$ and $\Xi'_Q$ is suppressed
both by $1/m$ and by $SU(3)$.
There is a large number of $P$-wave excited states;
we shall not discuss them here.

\begin{figure}[ht]
\begin{center}
\begin{picture}(90,60)
\put(50.5,33){\makebox(0,0){\includegraphics{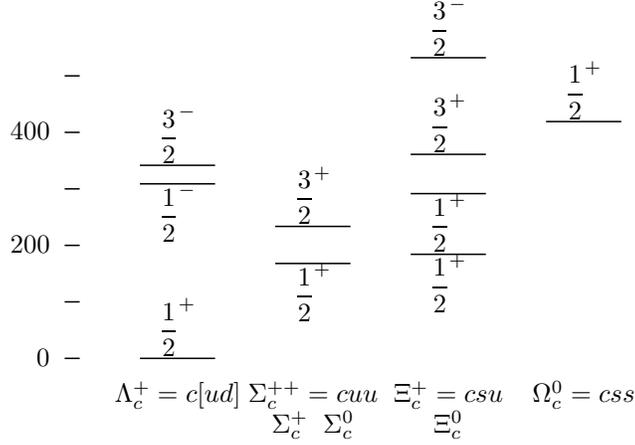}}}
\put(10,13){\makebox(0,0)[r]{0}}
\put(10,28.03){\makebox(0,0)[r]{200}}
\put(10,43.07){\makebox(0,0)[r]{400}}
\put(27,8){\makebox(0,0){$\Lambda_c^+=c[ud]$}}
\put(45,8){\makebox(0,0){$\Sigma_c^{++}=cuu$}}
\put(63,8){\makebox(0,0){$\Xi_c^+=csu$}}
\put(81,8){\makebox(0,0){$\Omega_c^0=css$}}
\put(45,4){\makebox(0,0){$\Sigma_c^+\hspace{2mm}\Sigma_c^0$}}
\put(63,4){\makebox(0,0){$\Xi_c^0$}}
\put(27,17){\makebox(0,0){$\displaystyle\frac{1}{2}^+$}}
\put(27,32.22){\makebox(0,0){$\displaystyle\frac{1}{2}^-$}}
\put(27,42.69){\makebox(0,0){$\displaystyle\frac{3}{2}^-$}}
\put(45,21.63){\makebox(0,0){$\displaystyle\frac{1}{2}^+$}}
\put(45,34.56){\makebox(0,0){$\displaystyle\frac{3}{2}^+$}}
\put(63,22.85){\makebox(0,0){$\displaystyle\frac{1}{2}^+$}}
\put(63,30.92){\makebox(0,0){$\displaystyle\frac{1}{2}^+$}}
\put(63,44.15){\makebox(0,0){$\displaystyle\frac{3}{2}^+$}}
\put(63,57){\makebox(0,0){$\displaystyle\frac{3}{2}^-$}}
\put(81,48.51){\makebox(0,0){$\displaystyle\frac{1}{2}^+$}}
\end{picture}
\end{center}
\caption{Baryons with $c$ quark}
\label{cbar}
\end{figure}

Experimentally observed~\cite{PDG:00} baryons containing $c$ quark
are shown in Fig.~\ref{cbar}.
The higher states in the first and the third columns are $P$-wave.
In the third column, the lowest state $\Xi_c$
is followed by the doublet $\Xi'_c$, $\Xi^*_c$.
The $\Omega^*_c$ baryon has not yet been observed.
The only baryon with $b$ quark, $\Lambda_b^0$,
has been discovered so far.

\section{Masses}

In the leading $m_b\to\infty$ approximation,
the masses $m_B$ and $m_{B^*}$ are both equal to $m_b+\bar{\Lambda}$,
where $\bar{\Lambda}$ is the energy of the ground state
of light fields in the chromoelectric field of the $\bar{b}$ antiquark.
This energy $\bar{\Lambda}$ is of order of $\Lambda_{\text{QCD}}$.
Excited states of light fields have energies $\bar{\Lambda}_i$,
giving excited degenerate doublets with the masses $m_b+\bar{\Lambda}_i$.

There are two $1/m_b$ corrections to the masses.
First, the $\bar{b}$ antiquark has an average momentum squared $\mu_\pi^2$,
which is of order of $\Lambda_{\text{QCD}}^2$.
Therefore, it has kinetic energy $\mu_\pi^2/(2m_b)$.
Second, the $\bar{b}$ chromomagnetic moment interacts with the chromomagnetic
field created by light constituents at the origin, where $\bar{b}$ stays.
This chromomagnetic field is proportional to
the light fields' angular momentum $\vec{j}_l$.
Therefore, the chromomagnetic interaction energy is proportional to
\begin{equation*}
\vec{s}_{Q}\cdot\vec{j}_l =
\frac{1}{2} \left[s(s+1) - s_{Q}(s_{Q}+1) - j_l(j_l+1)\right] =
\left\{
\begin{array}{ll}
\displaystyle         - \frac{3}{4}\,, & s=0 \\[2mm]
\displaystyle\phantom{-}\frac{1}{4}\,, & s=1
\end{array}
\right.
\end{equation*}
where $\vec{s}=\vec{s}_Q+\vec{j}_l$ is the meson spin.
If we denote this energy for $B$ as $-\mu_G^2/(2m_b)$,
then for $B^*$ it will be $\frac{1}{3}\mu_G^2/(2m_b)$.
Here $\mu_G^2$ is of order of $\Lambda_{\text{QCD}}^2$.
The $B$, $B^*$  meson masses with $1/m_b$ corrections taken into account
are given by the formulae
\begin{equation}
\begin{split}
m_B &{}= m_b + \bar{\Lambda} + \frac{\mu_\pi^2 - \mu_G^2}{2m_b}
+ \mathcal{O}\left(\frac{\Lambda_{\text{QCD}}^3}{m_b^2}\right)\,,\\
m_{B^*} &{}= m_b + \bar{\Lambda} + \frac{\mu_\pi^2 + \frac{1}{3}\mu_G^2}{2m_b}
+ \mathcal{O}\left(\frac{\Lambda_{\text{QCD}}^3}{m_b^2}\right)\,.
\end{split}
\label{mass}
\end{equation}

The hyperfine splitting is
\begin{equation*}
m_{B^*} - m_B = \frac{2\mu_G^2}{3m_b}
+ \mathcal{O}\left(\frac{\Lambda_{\text{QCD}}^3}{m_b^2}\right)\,.
\end{equation*}
Taking into account $m_{B^*}+m_B=2m_b+\mathcal{O}(\Lambda_{\text{QCD}})$,
we obtain
\begin{equation*}
m_{B^*}^2 - m_B^2 = \frac{4}{3}\mu_G^2
+ \mathcal{O}\left(\frac{\Lambda_{\text{QCD}}^3}{m_b}\right)\,.
\end{equation*}
The difference $m_{D^*}^2-m_D^2$ is given by a similar formula
with $m_c$ instead of $m_b$.
Therefore, the ratio
\begin{equation}
\frac{m_{B^*}^2-m_B^2}{m_{D^*}^2-m_D^2} = 1
+ \mathcal{O}\left(\frac{\Lambda_{\text{QCD}}}{m_{c,b}}\right)\,.
\label{ratio}
\end{equation}
Experimentally, this ratio is 0.89.
This is a spectacular confirmation of the idea
that violations of the heavy quark spin symmetry
are proportional to $1/m$.
In fact, the matrix element $\mu_G^2$ depends on the normalization scale,
and hence is not quite the same for $D$ and $B$;
this produces moderate perturbative corrections to~(\ref{ratio}).

\section{Strong decays}

$P$-wave excited states decay into the ground state emitting a pion.
In the ideal world with an infinitely heavy scalar $\bar{c}$,
the $\frac{1}{2}^+$ $P$-wave meson decays into the $\frac{1}{2}^-$
ground-state meson plus a pion (having $s^P=0^-$)
with the orbital angular momentum $l=0$ ($S$-wave);
the $\frac{3}{2}^+$ $P$-wave meson decays into the ground-state meson
plus a pion with $l=2$ ($D$-wave).
The pion momenta $p_\pi$ in these decays are rather small,
and hence the decay of the $\frac{3}{2}^+$ meson
(whose width is proportional to $p_\pi^5$) is strongly suppressed.
The decay of the $\frac{1}{2}^+$ meson is not suppressed,
its width is proportional to $p_\pi$.
Therefore, in our real world, the $0^+$, $1^+$ doublet mesons
are very wide, and difficult to observe.

Let's consider decays of $D_1$, $D_2$,
taking the $\bar{c}$ spin into account,
but still in the limit $m_c\to\infty$.
The widths of $D_1$ and $D_2$ are equal,
because the $\bar{c}$ spin plays no role in the decay.
$D_1$ decays only into $D^*\pi$, the decay into $D\pi$
is forbidden by the angular momentum conservation.
$D_2$ can decay into both $D\pi$ and $D^*\pi$.
In order to find the branching ratio $B(D_2\to D\pi)$,
we shall use a simple device known as Shmushkevich factory.

Let's take a sample of $D_{1,2}$ mesons
with random polarizations of $\bar{c}$.
Then 3/8 of this sample are $D_1$, and 5/8 are $D_2$.
Now let's wait for a small time $d t$;
a fraction $\Gamma dt$ of the sample will decay.
Ground-state mesons produced have randomly polarized $\bar{c}$.
Therefore, 1/4 of them are $D$, and 3/4 are $D^*$.
$D$ mesons are only produced from $D_2$:
\begin{equation*}
\frac{5}{8} B(D_2\to D\pi) = \frac{1}{4}\,,
\end{equation*}
and hence~\cite{IW:91c}
\begin{equation*}
B(D_2\to D\pi) = \frac{2}{5}\,,\quad
B(D_2\to D^*\pi) = \frac{3}{5}\,.
\end{equation*}

In the limit $m_c\to\infty$, $D$ and $D^*$ are degenerate,
and so are $D_1$ and $D_2$.
In the real world, pion momenta in these decays differ.
The widths are proportional to $p_\pi^5$,
and even rather small momentum differences produce a drastic effect.
It seems natural to suppose that the heavy-quark spin symmetry
predictions hold for the coefficients in front of $p_\pi^5$.
Then
\begin{equation*}
\frac{\Gamma(D_2\to D\pi)}{\Gamma(D_2\to D^*\pi)} =
\frac{2}{3} \left(\frac{p_\pi(D_2\to D\pi)}{p_\pi(D_2\to D^*\pi)}\right)^5 = 2.5\,,
\end{equation*}
while the experimental value is $2.3\pm0.6$~\cite{PDG:00}.
Formally, the difference of $p_\pi(D_2\to D\pi)/p_\pi(D_2\to D^*\pi)$
from 1 is a $1/m_c$ correction.
We can only hope that this kinematical $1/m_c$ effect,
included in the above estimate, is dominant.

\section{Leptonic decay constants}

Let's now discuss the $B$-meson leptonic decay constant $f_B$.
It is defined by
\begin{equation*}
{<}0|\bar{b}\gamma^\mu\gamma^5u|B(p){>} = i f_B p^\mu\,,
\end{equation*}
where the one-particle state is normalized in the usual Lorentz-invariant way:
\begin{equation*}
{<}B(p')|B(p){>} = 2 p^0 (2\pi)^3 \delta(\vec{p}'-\vec{p})\,.
\end{equation*}
This relativistic normalization becomes senseless in the limit $m_b\to\infty$,
and the non-relativistic normalization
\begin{equation*}
{}_{\text{nr}}{<}B(p')|B(p){>}_{\text{nr}} = (2\pi)^3 \delta(\vec{p}'-\vec{p})
\end{equation*}
should be used instead.
Then, for the $B$ meson at rest,
\begin{equation*}
{<}0|\bar{b}\gamma^0\gamma^5u|B{>}_{\text{nr}} = \frac{i m_B f_B}{\sqrt{2m_B}}\,.
\end{equation*}
Denoting this matrix element (which is mass-independent at $m_b\to\infty$)
as $i F/\sqrt{2}$, we obtain
\begin{equation}
f_B = \frac{F}{\sqrt{m_b}} \left[1 +
\mathcal{O}\left(\frac{\Lambda_{\text{QCD}}}{m_b}\right)\right]\,,
\label{fB}
\end{equation}
and hence
\begin{equation}
\frac{f_B}{f_D} = \sqrt{\frac{m_c}{m_b}} \left[1 +
\mathcal{O}\left(\frac{\Lambda_{\text{QCD}}}{m_{c,b}}\right)\right]\,.
\label{fRatio}
\end{equation}
In fact, the matrix element $F$ depends
on the normalization scale, and hence is not quite the same for $D$ and $B$;
this produces moderate perturbative corrections to~(\ref{fRatio}).
Lattice simulations and QCD sum rules show that the $1/m_c$ correction
in the formula for $f_D$ similar to~(\ref{fB}) is of order 100\%,
so that the accuracy of~(\ref{fRatio}) is not high.

Experimentally~\cite{PDG:00},
\begin{equation*}
f_{D_s^+}=280\pm19\pm28\pm34\,\text{MeV}\,,\quad
f_{D^+}=300_{-150-40}^{+180+80}\,\text{MeV}\,,
\end{equation*}
from the $\mu^+\nu_\mu$ and $\tau^+\nu_\tau$ decays.
The branching $B(B^+\to\tau^+\nu_\tau)$ should be of order $0.5\cdot10^{-4}$,
so that a direct measurement of $f_{B^+}$ at $B$-factories
seems feasible.
Theoretical estimates of $f_B$ vary by about a factor 2.

\section{Exclusive semileptonic decays}

Let's discuss the decay $B\to\bar{D}W^*$,
where $W^*$ is a virtual $W^+$ which decays into $l^+\nu_l$.
In the limit $m_b\to\infty$, $m_c\to\infty$,
it is enough to consider the case
when $\bar{b}$, $\bar{c}$, and $W^*$ are scalar.
We concentrate our attention on decays of $B$ with 4-velocity $v$
into $\bar{D}$ with 4-velocity $v'$.
Let $\J$ be the scalar current which replaces
a scalar $\bar{b}$ with 4-velocity $v$
by a scalar $\bar{c}$ with 4-velocity $v'$.
With the non-relativistic normalization
of the scalar quark wave functions, they are just 1,
and the quark decay matrix element is
\begin{equation}
{<}\bar{c}|\J|\bar{b}{>} = 1\,.
\label{Norm}
\end{equation}

The ground-state $B$ meson has $s^P=\frac{1}{2}^+$;
$\bar{D}$ will be used to denote generically
a ground-state or excited $\bar{c}q$ meson.
It is convenient to work in the $B$ rest frame.
Let the $z$ axis be in the direction of $\bar{D}$ motion.
Angular momentum conservation gives $s'_z=s_z$.
Reflection in a plane containing the $z$ axis
transforms a state $|s,s_z{>}$ into $P i^{2s}|s,-s_z{>}$.
Therefore, the amplitude of the $-s_z$ into $-s_z$ transition
is equal to that of the $s_z$ into $s_z$ transition,
up to a phase factor;
an $s_z=0$ into $s'_z=0$ transition is allowed
only when the ``naturalness'' $P(-1)^s$ is conserved~\cite{Po:90}.
For example,
the transition $\Lambda_b\to\Lambda_c$ is described by a single form factor;
$\Lambda_b\to\Sigma_c$ is forbidden by ``naturalness''
(and also suppressed by isospin);
$\Sigma_b\to\Sigma_c$ is described by two form factors
($s_z=s_z'=0$ and $\pm1$)~\cite{IW:91b,Ge:91,MRR:91}.

The transitions of the ground-state $\frac{1}{2}^+$ $\bar{b}q$ meson
into an $S$-wave $\frac{1}{2}^+$ $\bar{c}q$ meson,
and into a $P$-wave $\frac{1}{2}^-$ or $\frac{3}{2}^-$ $\bar{c}q$ meson,
are described by one form factor each~\cite{IW:90,FGGW:90,IW:91a,Fa:92}:
\begin{equation}
\begin{split}
{<}\tfrac{1}{2}^+|\J|\tfrac{1}{2}^+{>} &{}=
\xi(\cosh\vartheta) \bar{u}' u\,,\\
{<}\tfrac{1}{2}^-|\J|\tfrac{1}{2}^+{>} &{}=
\tau_{1/2}(\cosh\vartheta) \bar{u}'\gamma_5 u\,,\\
{<}\tfrac{3}{2}^-|\J|\tfrac{1}{2}^+{>} &{}=
\tau_{3/2}(\cosh\vartheta) v^\mu \bar{u}'_\mu u\,,
\end{split}
\label{IW}
\end{equation}
where $\cosh\vartheta=v\cdot v'$,
$\vartheta$ is the Minkowski angle
between the 4-velocities of $B$ and $\bar{D}$.

The Dirac wave function $u$ of the initial $\frac{1}{2}^+$ meson
satisfies $(\rlap/v-1)u=0$ and is normalized by the non-relativistic
condition $\bar{u}u=1$;
the sum over its two polarizations is
\begin{equation*}
\sum u \bar{u} = \frac{1+\rlap/v}{2}\,.
\end{equation*}
The Dirac wave function $u'$ of the final $\frac{1}{2}^\pm$ meson
has similar properties, with $v'$ instead of $v$.
The Rarita--Schwinger wave function $u'_\mu$ of the spin $\frac{3}{2}$ meson
satisfies $(\rlap/v'-1)u'_\mu=0$, $\gamma^\mu u'_\mu=0$ $v^{\prime\mu}u'_\mu=0$,
and is normalized by $\bar{u}^{\prime\mu}u'_\mu=-1$;
the sum over four polarization of the meson is
\begin{equation}
\sum u'_\mu \bar{u}'_\nu =
\frac{1+\rlap/v'}{2} \left( - g_{\mu\nu} + \frac{1}{3} \gamma_\mu \gamma_\nu
+ \frac{2}{3} v_\mu v_\nu \right) \frac{1+\rlap/v'}{2}\,.
\label{Rarita}
\end{equation}

\begin{figure}[ht]
\begin{center}
\begin{picture}(66,44)
\put(35,23){\makebox(0,0){\includegraphics{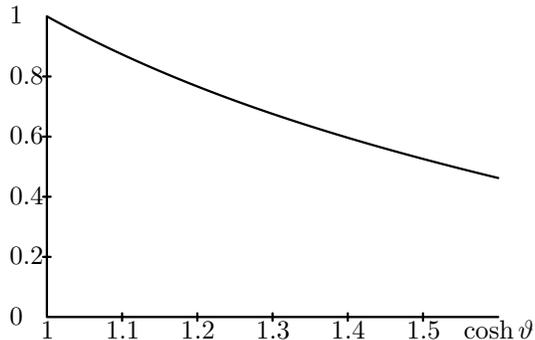}}}
\put(5,0){\makebox(0,0)[b]{$1$}}
\put(15,0){\makebox(0,0)[b]{$1.1$}}
\put(25,0){\makebox(0,0)[b]{$1.2$}}
\put(35,0){\makebox(0,0)[b]{$1.3$}}
\put(45,0){\makebox(0,0)[b]{$1.4$}}
\put(55,0){\makebox(0,0)[b]{$1.5$}}
\put(65,0){\makebox(0,0)[b]{$\cosh\vartheta$}}
\put(0,3){\makebox(0,0)[l]{$0$}}
\put(0,11){\makebox(0,0)[l]{$0.2$}}
\put(0,19){\makebox(0,0)[l]{$0.4$}}
\put(0,27){\makebox(0,0)[l]{$0.6$}}
\put(0,35){\makebox(0,0)[l]{$0.8$}}
\put(0,43){\makebox(0,0)[l]{$1$}}
\end{picture}
\end{center}
\caption{The Isgur--Wise form factor}
\label{IWexp}
\end{figure}

All the form factors of $B$ transitions into $\bar{D}$, $\bar{D}^*$
via the vector and axial $\bar{b}c$ weak currents
are proportional to the Isgur--Wise form factor $\xi(\cosh\vartheta)$,
with trivial kinematical coefficients.
When the current $J$ replaces an infinitely heavy $\bar{b}$
by an infinitely heavy $\bar{c}$ with the same 4-velocity and colour,
light fields don't notice it:
\begin{equation}
\xi(1) = 1\,.
\label{IWnorm}
\end{equation}
The maximum $\cosh\vartheta$ accessible
in the $B\to\bar{D}$, $\bar{D}^*$ decays is about 1.6;
a rough sketch of $\xi(\cosh\vartheta)$ as extracted from
experimental data is shown in Fig.~\ref{IWexp}.
At $\cosh\vartheta\gg1$, the Isgur--Wise form factor behaves as~\cite{GN:97}
\begin{equation}
\xi(\cosh\vartheta) \sim \frac{\as}{\cosh^2\vartheta}\,,
\label{asympt}
\end{equation}
up to logarithmic factors.

\begin{figure}[ht]
\begin{center}
\begin{picture}(42,22)
\put(21,11){\makebox(0,0){\includegraphics{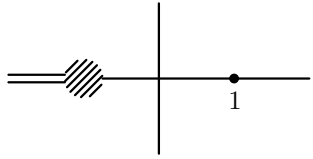}}}
\put(31,8){\makebox(0,0){1}}
\end{picture}
\end{center}
\caption{The complex $\cosh\vartheta$ plane}
\label{IWcomplex}
\end{figure}

The $B\to B$ form factor is also proportional to $\xi(\cosh\vartheta)$.
In this case, $q^2=2m_B^2(1-\cosh\vartheta)$.
The form factor has a cut in the annihilation channel
from $q^2=4m_B^2$ to $+\infty$.
Therefore, $\xi(\cosh\vartheta)$ has a cut
from $\cosh\vartheta=-1$ to $-\infty$ (Fig.~\ref{IWcomplex}).
Geometrically speaking, $\cosh\vartheta>1$ corresponds to
Minkowski angles between the world lines of the incoming heavy quark
and the outgoing one -- this is the scattering (or decay) channel.
When $\cosh\vartheta=1$, the world line is straight,
and there is no transition at all (see~(\ref{IWnorm})).
When $|\cosh\vartheta|<1$, the angle is Euclidean.
When $\cosh\vartheta<-1$, we have a Minkowski angle again,
but one of the 4-velocities is directed into the past ---
this is the annihilation channel.
At the point $\cosh\vartheta=-1$, the heavy quark returns back
along the same world line.
In fact, the very concept of the Isgur--Wise form factor
is inapplicable near this point.
The HQET picture is based on the fact that
heavy quarks move along straight world lines.
If their relative velocity in the annihilation channel
is $\lesssim\as$, they rotate around each other instead.
The $B$ meson form factor has poles below the threshold
corresponding to $\Upsilon$ mesons
with binding energies $\sim m_b\as^2$;
its behaviour in this region is not universal.
The concept of the Isgur--Wise form factor is only applicable
at $|\cosh\vartheta+1|\gg\as^2$ (Fig.~\ref{IWcomplex}).

Squaring the matrix elements~(\ref{IW}),
summing over the final meson polarizations
(using~(\ref{Rarita}) for the sipn $\frac{3}{2}$ meson),
averaging over the initial meson polarizations,
and normalizing to the quark decay~(\ref{Norm}),
we obtain the branching ratios
\begin{equation}
\begin{split}
B\left(\tfrac{1}{2}^+\to\tfrac{1}{2}^+\right) &{}=
\frac{\cosh\vartheta+1}{2} \xi^2(\cosh\vartheta)\,,\\
B\left(\tfrac{1}{2}^+\to\tfrac{1}{2}^-\right) &{}=
\frac{\cosh\vartheta-1}{2} \tau_{1/2}^2(\cosh\vartheta)\,,\\
B\left(\tfrac{1}{2}^+\to\tfrac{3}{2}^-\right) &{}=
\frac{(\cosh\vartheta+1)^2 (\cosh\vartheta-1)}{3}
\tau_{3/2}^2(\cosh\vartheta)\,.
\end{split}
\label{Br}
\end{equation}
They are the fractions of the number of $B\to X_{\bar{c}}W^+$ decays
with $X_{\bar{c}}$ velocity $v'$,
where the hadronic system $X_{\bar{c}}$ happens to be a single meson.
The decay $\frac{1}{2}^+\to\frac{1}{2}^+$ is $S$-wave,
and therefore the squared matrix element tends to a constant
at $\vartheta\to0$.
The decays $\frac{1}{2}^+\to\frac{1}{2}^-$
and $\frac{1}{2}^+\to\frac{3}{2}^-$ are $P$-wave, 
and therefore the squared matrix elements behave
as the relative velocity squared, $\vartheta^2$.
Similarly, decays into $D$-wave mesons $\frac{3}{2}^+$, $\frac{5}{2}^+$
are $D$-wave, and behave as $\vartheta^4$ at $\vartheta\to0$.
If we choose the mass of the virtual $W^*$ larger than $m_B+m_D$,
we can consider the channel $W^*\to BD$.
The squared matrix elements are given by the formulae~(\ref{Br})
with the extra factor 2, because we sum over the $D$ polarizations now,
not average.
The decays of the scalar $W^*$ into $\frac{1}{2}^+\frac{1}{2}^-$,
$\frac{1}{2}^+\frac{1}{2}^+$, and $\frac{1}{2}^+\frac{3}{2}^+$
are $P$-wave, $S$-wave, and $D$-wave.
Therefore, their squared matrix elements behave as the relative velocity
to the power 2, 0, and 4, correspondingly.
This explains the behaviour of the formulae~(\ref{Br})
at $\cosh\vartheta\to-1$.

\section{Inclusive semileptonic decays}

\begin{figure}[ht]
\begin{center}
\begin{picture}(102,64)
\put(51,33){\makebox(0,0){\includegraphics{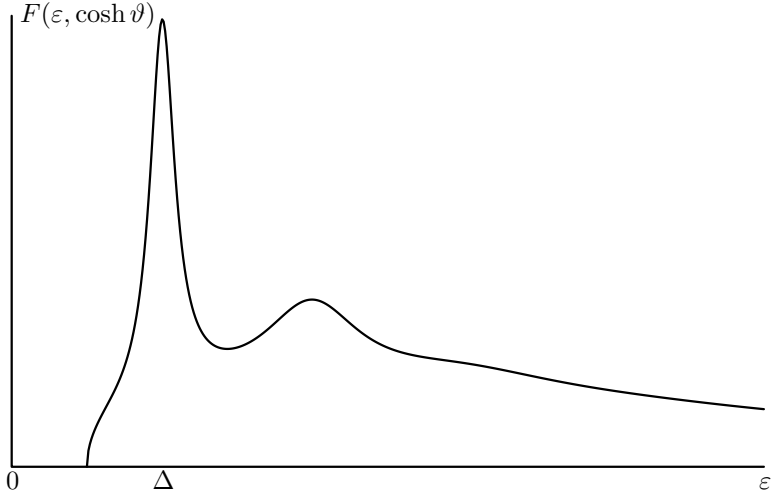}}}
\put(1,0){\makebox(0,0)[b]{$0$}}
\put(21,0){\makebox(0,0)[b]{$\Delta$}}
\put(101,0){\makebox(0,0)[b]{$\varepsilon$}}
\put(2,63){\makebox(0,0)[l]{$F(\varepsilon,\cosh\vartheta)$}}
\end{picture}
\end{center}
\caption{A qualitative sketch of the $B\to X_{\bar{c}}$ decay structure function}
\label{struct}
\end{figure}

The inclusive decay rate $B\to X_{\bar{c}}W^*$ can be written as
$F(\varepsilon,\cosh\vartheta)d\varepsilon$,
where $\cosh\vartheta=v\cdot v'$,
$v'$ is the $X_{\bar{c}}$ 4-velocity: $p_X=m_X v'$,
and $\varepsilon=m_X-m_D$ is the excitation energy
(we are still in the limit $m_c\to\infty$, where $m_{D^*}=m_D$).
The structure function is
\begin{align}
F(\varepsilon,\cosh\vartheta) ={}&
\frac{\cosh\vartheta+1}{2}
\sum_{\frac{1}{2}^+} \xi_i^2(\cosh\vartheta) \delta(\varepsilon-\varepsilon_i)
\nonumber\\
&{} + \frac{\cosh\vartheta-1}{2}
\sum_{\frac{1}{2}^-} \tau_{1/2}^2(\cosh\vartheta) \delta(\varepsilon-\varepsilon_i)
\label{Str}\\
&{} + \frac{(\cosh\vartheta+1)^2 (\cosh\vartheta-1)}{3}
\sum_{\frac{3}{2}^-} \tau_{3/2}^2(\cosh\vartheta) \delta(\varepsilon-\varepsilon_i)
+ \cdots
\nonumber
\end{align}
where the sums run over final states with the indicated quantum numbers,
$\varepsilon_i$ are their excitation energies,
the index $i$ is not explicitly shown
in the form factors $\tau_{1/2}$ and $\tau_{3/2}$,
and the dots mean the contribution of $D$-wave and higher states.
At $\vartheta=0$, $F(\varepsilon,1)=\delta(\varepsilon)$.
A qualitative sketch of $F(\varepsilon,\cosh\vartheta)$
as a function of $\varepsilon$ at some fixed $\vartheta>0$
is shown in Fig.~\ref{struct}.
It contains a $\delta$ peak at $\varepsilon=0$ due to the transition
into the ground state ($\bar{D}$ and $\bar{D}^*$),
then some peaks due to excited states which become wider
when $\varepsilon$ increases, and the curve becomes smooth.
At $\varepsilon\gg\Lambda_{\text{QCD}}$,
it is given by the perturbative gluon radiation:
\begin{equation}
F(\varepsilon,\cosh\vartheta) =
\frac{2 C_F \as (\vartheta\coth\vartheta-1)}{\pi\varepsilon}\,.
\label{pert}
\end{equation}
Its $\varepsilon$ dependence is evident from dimensionality,
and the $\vartheta$ dependence is given by the famous QED
soft-photon radiation function.
It is also known from classical electrodynamics:
this function is the distribution in the radiation energy
when a charge suddenly changes its velocity from $v$ to $v'$.

The total decay probability is unity:
\begin{equation}
\int F(\varepsilon,\cosh\vartheta) d\varepsilon = 1\,.
\label{Bj}
\end{equation}
This is the Bjorken sum rule~\cite{Bj:90,BDT:92,IW:91a}.
In particular, the decay rate into the ground-state $\frac{1}{2}^+$ meson
must not exceed the total one:
\begin{equation}
\xi(\cosh\vartheta) \le \sqrt{\frac{2}{\cosh\vartheta+1}}\,.
\label{Bj1}
\end{equation}
At $\cosh\vartheta\gg1$, the ground state is rarely produced,
and the Isgur--Wise form factor~(\ref{asympt}) is much less
than the bound~(\ref{Bj1}).

The Bjorken sum rule becomes much simpler in the small $\vartheta$ limit.
The Isgur--Wise form factor of the transition into the ground state
behaves as $\xi(\cosh\vartheta)=1-\rho^2(\cosh\vartheta-1)+\cdots$,
and those of transitions into higher $S$-wave mesons -- as
$\xi_i(\cosh\vartheta)=\rho_i^2(\cosh\vartheta-1)+\cdots$.
Expanding~(\ref{Bj}) up to linear terms in $\cosh\vartheta-1$,
we obtain
\begin{equation}
1 + \left[\frac{1}{2} - 2\rho^2
+ \frac{1}{2} \sum \tau_{1/2}^2(1) + \frac{4}{3} \sum \tau_{3/2}^2(1)
\right] (\cosh\vartheta-1) = 1\,.
\label{Bj2}
\end{equation}
$D$-wave final state don't contribute in this order,
as well as higher $S$-wave final states.
Therefore, the slope of the Isgur--Wise form factor $\rho^2$
is expressed via the form factors of $P$-wave meson production
at $\cosh\vartheta=1$:
\begin{equation}
\rho^2 = \frac{1}{4} + \frac{1}{4} \sum \tau_{1/2}^2(1)
+ \frac{2}{3} \sum \tau_{3/2}^2(1)\,.
\label{Bj3}
\end{equation}
In particular,
\begin{equation}
\rho^2 > \frac{1}{4}
\label{Bj4}
\end{equation}
(this also follows from~(\ref{Bj1})).

We can also consider the inclusive decay of polarized $B$.
Its structure function is
(we use~(\ref{Rarita}) for the spin $\frac{3}{2}$ meson)
\begin{align*}
&\bar{u} \frac{\rlap/v'+1}{2} u
\sum_{\frac{1}{2}^+} \xi_i^2(\cosh\vartheta) \delta(\varepsilon-\varepsilon_i)
+ \bar{u} \frac{\rlap/v'-1}{2} u
\sum_{\frac{1}{2}^-} \tau_{1/2}^2(\cosh\vartheta) \delta(\varepsilon-\varepsilon_i)\\
&{} + \frac{\cosh\vartheta+1}{3}
\left[ 2\cosh\vartheta-1 - (2-\cosh\vartheta) \bar{u}\rlap/v'u \right]
\sum_{\frac{3}{2}^-} \tau_{3/2}^2(\cosh\vartheta) \delta(\varepsilon-\varepsilon_i)\\
&{} + \cdots
\end{align*}
Averaging it over polarizations $\bar{u}\rlap/v'u\to\cosh\vartheta$,
we reproduce~(\ref{Str}).
The decay rate does not depend on the initial meson polarization.
This gives the Uraltsev sum rule~\cite{Ur:01}
\begin{equation}
\begin{split}
\sum_{\frac{1}{2}^+} \xi_i^2(\cosh\vartheta)
&{}+ \sum_{\frac{1}{2}^-} \tau_{1/2}^2(\cosh\vartheta)\\
&{}- \frac{2}{3} (\cosh\vartheta+1) (2-\cosh\vartheta)
\sum_{\frac{3}{2}^-} \tau_{3/2}^2(\cosh\vartheta)
+ \cdots = 0\,.
\end{split}
\label{Ur1}
\end{equation}
It becomes much simpler at $\vartheta\to0$.
$D$- and higher-wave contributions vanish, and
\begin{equation}
1 + \sum \tau_{1/2}^2(1) - \frac{4}{3} \sum \tau_{3/2}^2(1) = 0\,.
\label{Ur2}
\end{equation}
Substituting $\sum\tau_{3/2}^2(1)$ from the this sum rule into~(\ref{Bj3}),
we obtain
\begin{equation}
\rho^2 = \frac{3}{4} \left( 1 + \sum \tau_{1/2}^2(1) \right)\,.
\label{Ur3}
\end{equation}
In particular,
\begin{equation}
\rho^2 > \frac{3}{4}\,.
\label{Ur4}
\end{equation}
This Uraltsev boundary is much stronger than the Bjorken boundary~(\ref{Bj4}).
Experimantally, $\rho^2\sim0.8$.

More sum rules can be obtained from the energy conservation.
In the $v$ rest frame, the light fields in $B$
have a definite energy $E=\bar{\Lambda}$.
They have no definite momentum, because they are in the external chromoelectric field
created by the $\bar{b}$ antiquark.
Its average is ${<}\vec{p}{>}=0$,
and the average of its square is ${<}\vec{p}^2{>}=\mu_\pi^2$
(it is the same as the average squared momentum of the heavy antiquark,
see~(\ref{mass})).
The light fields' energy in the $v'$ frame is
$E'=\bar{\Lambda}\cosh\vartheta-p_x\sinh\vartheta$.
When $\bar{b}$ with the 4-velocity $v$ is suddenly transformed
into $\bar{c}$ with the 4-velocity $v'$,
the light fields remain in their original state at the first moment.
After that, the energy $E'$ is conserved in the field of $\bar{c}$
moving with the 4-velocity $v'$.
Therefore, the average excitation energy of $X_{\bar{c}}$
and the average squared excitation energy are
\begin{align*}
{<}E'-\bar{\Lambda}{>} &{}= \bar{\Lambda}(\cosh\vartheta-1),\\
{<}(E'-\bar{\Lambda})^2{>} &{}= \bar{\Lambda}^2(\cosh\vartheta-1)^2
+ \frac{\mu_\pi^2}{3}(\cosh^2\vartheta-1)
\end{align*}
(because ${<}p_x^2{>}={<}\vec{p}^2{>}/3$).
This gives the Voloshin sum rule~\cite{Vo:92}
and the BGSUV sum rule~\cite{BGSUV:94}:
\begin{align}
\int F(\varepsilon,\cosh\vartheta) \varepsilon d\varepsilon &{}=
\bar{\Lambda}(\cosh\vartheta-1)\,,
\label{Vol}\\
\int F(\varepsilon,\cosh\vartheta) \varepsilon^2 d\varepsilon &{}=
\bar{\Lambda}^2(\cosh\vartheta-1)^2 + \frac{\mu_\pi^2}{3}(\cosh^2\vartheta-1)\,.
\label{BGSUV}
\end{align}
The transition into the ground-state meson with $\varepsilon=0$
does not contribute here.

Expanding these sum rules up to linear terms in $\cosh\vartheta-1$,
we obtain, similarly to~(\ref{Bj2}),
\begin{align}
\frac{1}{2} \sum \tau_{1/2}^2(1) \varepsilon_i
+ \frac{4}{3} \sum \tau_{3/2}^2(1) \varepsilon_i
&{}= \bar{\Lambda}\,,
\label{Vol2}\\
\frac{1}{2} \sum \tau_{1/2}^2(1) \varepsilon_i^2
+ \frac{4}{3} \sum \tau_{3/2}^2(1) \varepsilon_i^2
&{}= \frac{2}{3}\mu_\pi^2\,.
\label{BGSUV2}
\end{align}
Let $\Delta$ be the minimum $P$-wave excitation energy.
Then, replacing $\varepsilon_i$ by $\Delta$
in the left-hand side of~(\ref{Vol2}), we make it smaller.
After singling out this factor $\Delta$,
the remaining sum is $2\left(\rho^2-\frac{1}{4}\right)$~(\ref{Bj3}):
\begin{equation}
\bar{\Lambda} \ge 2 \Delta \left(\rho^2-\frac{1}{4}\right)\,.
\label{Vol3}
\end{equation}
This inequality can also be rewritten
as an upper bound on $\rho^2$~\cite{Vo:92}:
\begin{equation}
\rho^2 \le \frac{1}{4} + \frac{\bar{\Lambda}}{2\Delta}\,.
\label{Vol4}
\end{equation}

Similarly, replacing $\varepsilon_i^2$ by $\Delta\varepsilon_i$
in the left-hand side of~(\ref{BGSUV2}), we make it smaller.
After singling out the factor $\Delta$,
the remaining sum is $\bar{\Lambda}$~(\ref{Vol2}),
and~\cite{BGSUV:94}
\begin{equation}
\mu_\pi^2 \ge \frac{3}{2} \Delta \bar{\Lambda}
\ge 3 \Delta^2 \left(\rho^2-\frac{1}{4}\right)
\label{BGSUV3}
\end{equation}
(at the second step, the inequality~(\ref{Vol3}) was used).
If the lowest resonances in the $\frac{1}{2}^-$ and $\frac{3}{2}^-$ channels
with nearly equal energies dominate in~(\ref{Vol2}), (\ref{BGSUV2}),
then the inequalities~(\ref{Vol3}), (\ref{BGSUV3})
should be close to equalities.
This is, probably, the case.
Strictly speaking, the minimum excitation energy $\Delta$ is equal to $m_\pi$,
because the ground-state meson plus a soft pion
can have the needed quantum numbers.
This $\Delta$ is small, and the bounds are very weak.
However, the coupling of a soft pion with a heavy meson is small,
and these states contribute little
to the sum rules~(\ref{Vol2}), (\ref{BGSUV2}).
The first important contribution comes from the lowest $P$-wave resonances.

As you may have noticed, the integral in the Bjorken sum rule~(\ref{Bj})
logarithmically diverges at large $\varepsilon$, due to~(\ref{pert}).
The integrals~(\ref{Vol}), (\ref{BGSUV}) diverge even more strongly.
Therefore, if we want to take $\as$ effects into account,
we have to cut these integrals somehow.
On the other hand, the form factors $\xi_i(\cosh\vartheta)$,
$\tau_{1/2}(\cosh\vartheta)$, $\tau_{3/2}(\cosh\vartheta)$
depend on the normalization point $\mu$,
if perturbative effects are taken into consideration.
It is natural to expect that the sum rules are valid,
up to $\mathcal{O}(\as(\mu))$ corrections,
when $\mu$ is of the order of the cutoff energy
(this is discussed in~\cite{GK:96} in more details).
Therefore, the anomalous dimension of the HQET heavy-heavy quark current
(which describes the $\mu$-dependence of the formfactors)
is proportional to the soft-photon radiation function~(\ref{pert}).

For more information about inclusive sum rules,
see~\cite{BSUV:95,BSU:97,Ur:98,YOR:02}.

\section{Pair production of heavy mesons}

It is also possible to establish a bound on the Isgur--Wise form factor
at the cut (Fig.~\ref{IWcomplex}).
The decay rate $W^*\to BD$ must be less
than the total decay rate $W^*\to\bar{b}c$:
\begin{equation}
n_l |\xi(\cosh\vartheta)|^2 |\cosh\vartheta+1| \le N_c\,,
\label{dRT}
\end{equation}
where $n_l$ is the number of light flavours,
and $N_c$ is the number of colours.
The left-hand side should be the sum over light flavours,
if $SU(3)$ breaking is taken into account.
The inequality~(\ref{dRT}) is applicable
at $|\cosh\vartheta+1|\gg\as^2$
(outside the pitched region in Fig.~\ref{IWcomplex}),
where the Isgur--Wise form factor has sense,
and the decay rate $W^*\to\bar{b}c$ is given by the free quark formula.
At $|\cosh\vartheta|\gg1$,
production of the pair of ground-state mesons is rare,
and the Isgur--Wise form factor~(\ref{asympt})
is much less than the bound~(\ref{dRT}).
The inequality~(\ref{dRT}) was proposed in~\cite{RT:92},
where the factor $n_l$ has been erroneously omitted.
It was used, together with analyticity,
%and some information about form factor poles below threshold,
to obtain bounds on the Isgur--Wise form factor in the physical region.

\end{document}